# Treatment Choice in Heterogeneous Populations Using Experiments without Covariate Data

**Charles F. Manski**
Department of Economics and Institute for Policy Research
Northwestern University
Evanston, IL 60208

## Abstract

I examine the problem of treatment choice when a planner observes (i) covariates that describe each member of a population of interest and (ii) the outcomes of an experiment in which subjects randomly drawn from this population are randomly assigned to treatment groups within which all subjects receive the same treatment. Covariate data for the subjects of the experiment are not available. The optimal treatment rule is to divide the population into subpopulations whose members share the same covariate value, and then to choose for each subpopulation a treatment that maximizes its mean outcome. However the planner cannot implement this rule. I draw on my work on nonparametric analysis of treatment response to address the planner's problem.

## 1. INTRODUCTION

Suppose that a planner must choose a treatment rule assigning a treatment to each member of a heterogeneous population. Each person in the population has a response function mapping treatments into a real-valued outcome of interest. The planner wants to choose a treatment rule that maximizes the population mean outcome.

The feasible treatment rules and the planner's ability to evaluate them depend on the information that the planner possesses about the members of the population. If the planner somehow observes or otherwise knows each person's response function, he can choose treatments that maximize each person's outcome. However response functions generally are neither completely observed nor a priori known. Hence we are concerned with situations in which less extensive information is available.

This paper considers the problem of treatment choice when the planner has two types of empirical information about the population of interest. He observes (i) covariates that describe each member of the population and (ii) the outcomes of an experiment in which randomly drawn subjects are randomly assigned to treatment groups within which all subjects receive the same treatment. Covariate data for the subjects of the experiment are not available. Moreover, the planner has no observational (i.e., non-experimental) data on treatment response.

This informational situation is common in medical and other settings when new treatments are approved for use following a period of experimentation. Consider, for example, a physician who must choose treatments for a population of heterogeneous patients. Physicians commonly have extensive covariate information — medical histories, diagnostic test findings, and demographic data — for the patients that they treat. Physicians often know the outcomes of randomized clinical trials evaluating new treatments. The medical journal articles that report the findings of clinical trials, however, do not usually report extensive covariate information for the subjects of the experiment. Medical researchers reporting on clinical trials usually describe outcomes only within broad risk-factor groups.

A planner in the informational situation just described faces an interesting predicament. The optimal treatment rule given the observed covariates is to divide the population into subpopulations whose members share the same covariate value, and then to choose for each subpopulation a treatment that maximizes its mean outcome (see Section 2.2). However the available experimental data only reveal mean outcomes in the population as a whole — the planner does not observe mean outcomes in each subpopulation.

To address this problem, I draw on elements of my work on nonparametric analysis of treatment response. The methodological research program initiated in Manski (1989, 1990) and carried forward in Manski (1993, 1994, 1995, 1996, 1997a, 1997b) moves away from the longstanding focus of social scientists on exact identification of treatment effects. I have found that informative bounds on treatment effects may be obtained under various weak nonparametric assumptions. Recently, in Manski (1998), I have begun to examine the implications of these findings for treatment choice. The present paper develops these implications in a specific setting of interest.

Section 2 draws on Manski (1998) to develop general themes about the use of covariates in treatment choice. I show that an optimal treatment rule assigns to each member of the population a treatment that maximizes mean outcome conditional on the person's observed covariates. The planner is said to face a problem of treatment choice under *uncertainty* if he knows the conditional mean responses and, consequently, can implement an optimal rule. The planner is said to face a problem of treatment choice under *ambiguity* if he does not know enough about mean response to be able to implement an optimal rule.

Section 3 examines the particular problem of treatment choice under ambiguity that arises when the planner observes experimental outcomes without accompanying covariate data. I focus on the simplest non-trivial case; that in which treatments, outcomes, and covariates are all binary. There are four feasible treatment rules in this setting; two rules assign the same treatment to all persons and two assign different treatments to persons with different covariate values. I apply a finding of Manski (1997a) to determine which of the four rules are dominated and, consequently, should not be chosen by the planner. It turns out that the dominated rules depends on a somewhat subtle interplay of the distribution of covariates in the population and the distributions of outcomes revealed by the randomized experiment. There are as many as three or as few as zero dominated treatment rules, depending on the configuration of these distributions.

An empirical illustration helps to see the range of possibilities. Section 4 considers the situation of a planner, perhaps a social worker, who is charged with making preschool treatment choices for a population of children. The planner observes a binary covariate for each child. The planner observes the outcomes of a famous randomized experiment, the Perry Preschool Project, that assigned children to alternative preschool treatments. Application of the findings of Section 3 shows how the planner's ability to rank alternative treatment rules depends on the distribution of covariates in the population.

Section 5 gives conclusions.

# 2. USING COVARIATES TO CHOOSE TREATMENTS

## 2.1. THE PLANNER'S CHOICE SET AND OBJECTIVE FUNCTION

To formalize the problem of treatment choice, I suppose that the planner must choose a treatment rule assigning a treatment to each member of a specified population J. Each person $j \in J$ has an individual-specific *response function* $y_j(\cdot): T \to Y$ mapping the mutually exclusive and exhaustive *treatments* $t \in T$ into real-valued *outcomes* $y_j(t) \in Y$. A *treatment rule* is a function $\tau(\cdot): J \to T$ specifying which treatment each person receives; thus person j's outcome under rule $\tau(\cdot)$ is $y_j[\tau(j)]$.

The planner is concerned with the distribution of outcomes across the population, not with the experiences of particular individuals. With this in mind, I take the population to be a probability space, say $(J, \Omega, P)$, where $\Omega$ is the $\sigma$-algebra on which probabilities are defined and P is the probability measure. Now the population mean outcome under treatment rule $\tau(\cdot)$ is

$$(1) \quad E\{y_j[\tau(j)]\} \equiv \int y_j[\tau(j)]dP.$$

I assume that the planner wants to choose a treatment rule that maximizes $E\{y_j[\tau(j)]\}$. This criterion function has both normative and analytical appeal. Maximization of a population mean outcome, or perhaps some weighted average outcome, is the standard utilitarian criterion of the public economics literature on social planning. The linearity of the expectation operator yields substantial analytical simplifications, particularly through use of the law of iterated expectations.

I suppose that the planner observes certain covariates $x_j \in X$ for each member of the population. The planner cannot distinguish among persons with the same observed covariates. Hence he cannot implement treatment rules that systematically differentiate among these persons. With this in mind, I take the feasible rules be the set of functions mapping the observed covariates into treatments.[1]

---

[1] Although the planner cannot systematically differentiate among persons with the same observed covariates, he can randomly assign different treatments to such persons. Thus the set of feasible treatment rules in principle contains not only functions mapping covariates into treatments but also probability mixtures of these functions. Explicit consideration of randomized treatment rules would not substantively change the analysis of this paper, but would complicate the necessary notation. A simple implicit way to permit randomized rules is to include in x a component whose value is randomly drawn by the planner from some distribution. The planner can then make the chosen treatment vary with this covariate component.

To formalize this, let Z denote the space of all functions mapping X into T. Then the feasible rules have the form

$$(2) \quad \tau(j) = z(x_j), \qquad j \in J,$$

where $z(\cdot) \in Z$. The planner wants to solve the problem

$$(3) \quad \max_{z(\cdot) \in Z} \; E\{y[z(x)]\}.$$

## 2.2. OPTIMAL TREATMENT CHOICE UNDER UNCERTAINTY

It is easy to show that the optimal choice among the set Z of feasible treatment rules assigns to each member of the population a treatment that maximizes mean outcome conditional on the person's observed covariates. For each $z(\cdot) \in Z$, use the law of iterated expectations to write

$$(4) \quad E\{y[z(x)]\} = E\{E\{y[z(x)] \,|\, x\}\}$$
$$= E\{ \sum_{t \in T} E[y(t) \,|\, x] \cdot 1[z(x) = t]\}.$$

For each $x \in X$, the bracketed expression on the right side is maximized by choosing $z(x)$ to maximize $E[y(t) \,|\, x]$ on $t \in T$. Hence a treatment rule $z^*(\cdot)$ is optimal if

$$(5) \quad z^*(x) = \underset{t \in T}{\mathrm{argmax}} \; E[y(t) \,|\, x], \qquad x \in X.$$

The optimized population mean outcome is

$$(6) \quad V^* = E\{ \max_{t \in T} E[y(t) \,|\, x]\}.$$

The planner is said to face a problem of treatment choice under *uncertainty* if he knows the conditional mean responses $E[y(\cdot) \,|\, x]$, $x \in X$ and, consequently, can implement the optimal treatment rule.[2] It is easy to show that in a problem of treatment under uncertainty, observation of the covariates x cannot decrease the population mean outcome achieved by the planner. Suppose that the planner were to observe only w(x), where $w(\cdot): X \to W$ is a many-to-one function of x. Then the feasible treatment rules would be those that give the same treatment to each person with covariates w(x). These rules yield mean outcomes $E[y(t) \,|\, w(x)]$, $t \in T$; hence the optimized mean outcome is $E\{\max_{t \in T} E[y(t) \,|\, w(x)]\}$. With x observed, the optimized mean outcome is given in (6). By Jensen's Inequality,

$$(7) \quad E\{\max_{t \in T} E[y(t) \,|\, x]\} \; \geq \; E\{\max_{t \in T} E[y(t) \,|\, w(x)]\}.$$

Hence the optimized mean outcome using covariates x to choose treatments is necessarily at least as large as that achievable using the more limited covariates w(x).

## 2.3. UNDOMINATED TREATMENT CHOICE UNDER AMBIGUITY

The planner is said to face a problem of treatment choice under *ambiguity* if he does not know enough about the mean response functions $E[y(\cdot) \,|\, x]$, $x \in X$ to be able to implement the optimal treatment rule (5).

Suppose the planner knows only that the population (covariate, response function) distribution $P[x, y(\cdot)]$ lies within a specified set $\Phi$ of possible (covariate, response function) distributions. The planner may then partition the feasible treatment rules into dominated and undominated subclasses. A feasible treatment rule $z(\cdot)$ is dominated if there exists another feasible rule, say $z'(\cdot)$, such that

$$(8a) \quad \int y[z(x)]d\phi \; \leq \; \int y[z'(x)]d\phi, \quad \text{all } \phi \in \Phi,$$

$$(8b) \quad \int y[z(x)]d\phi \; < \; \int y[z'(x)]d\phi, \quad \text{some } \phi \in \Phi.$$

A treatment rule $z(\cdot)$ is undominated if no such $z'(\cdot)$ exists. A planner facing a problem of treatment under ambiguity can eliminate dominated rules as sub-optimal but cannot rank the undominated rules.[3]

---

[2]   The planner faces a problem of treatment choice under *certainty* if the outcome distributions $P[y(\cdot) \,|\, x]$, $x \in X$ are all degenerate. In this case, knowing $E[y(\cdot) \,|\, x]$, $x \in X$ means knowing each person's response function, with probability one.

---

[3] Bayesian decision theorists suggest using subjective outcome distributions to rank undominated rules. Bayesians offer various procedural rationality arguments for this approach. These arguments do not, however, answer the question most relevant to the planner: How well does the rule perform? See Manski (1998, Section 2.1) for further discussion.

A generic feature of decision problems under ambiguity is that expansion of the choice set may decrease welfare. In the treatment-choice setting, the choice set is the space of functions mapping covariates into treatments, so observation of additional covariates implies expansion of the choice set. Observation of additional covariates enables the planner to choose a treatment rule that more finely differentiates among the members of the population. The problem is that the planner, not knowing mean outcomes conditional on these covariates, may unwittingly use them to choose a worse treatment rule.

To see this, it suffices to consider the extreme case where the planner has no knowledge of $P[x, y(\cdot)]$, so $\Phi$ is the set of all (covariate, response function) distributions. In this case all feasible treatment rules are undominated and the planner may inadvertently choose the worst possible one. If only $w(x)$ is observed, the feasible treatment rules give the same treatment to every person with covariates $w(x)$. Then the worst possible rule yields $E\{\min_{t \in T} E[y(t)|w(x)]\}$ as the population mean outcome. If $x$ is observed, the feasible treatment rules give the same treatment to every person with covariates $x$. Now the worst possible rule yields $E\{\min_{t \in T} E[y(t)|x]\}$ as the population mean outcome. By Jensen's Inequality,

(9)   $E\{\min_{t \in T} E[y(t)|x]\} \leq E\{\min_{t \in T} E[y(t)|w(x)]\}$.

Hence using $x$ to choose treatments may decrease the population mean outcome achieved by the planner.

## 3.    TREATMENT CHOICE WHEN TREATMENTS, OUTCOMES, AND COVARIATES ARE BINARY

### 3.1. THE PLANNER'S INFORMATION AND THE FEASIBLE TREATMENT RULES

I now formalize the informational situation described in the Introduction. I assume that the planner has no prior information about treatment response; hence treatment choice must be based on the available empirical evidence alone. The planner observes the covariates $x$ for each member of the population and therefore knows the covariate distribution $P(x)$. The planner has no observational data on treatment response. He observes the outcomes of an experiment conducted on the population of interest, but does not observe covariates for the experimental subjects. I assume that the experiment achieves the classical ideal — the subjects are randomly drawn from the population and are randomly assigned to treatments, all subjects comply with their treatments, there are no interactions among subjects, and so on.   To keep attention focused on the identification problem that is at the heart of the planner's predicament, I assume that the samples of subjects are large enough that the planner may safely abstract from sampling variability when interpreting the empirical evidence. Then the experimental data reveal the outcome distributions $P[y(t)]$, $t \in T$.

The planner's problem is to use the available empirical evidence to make treatment choices. A central feature of treatment choice under ambiguity is that the normative question "What treatment rule should the planner choose?" has no clear answer. However the normative question "What treatment rules should the planner not choose?" does have a clear answer. That is, the planner should not choose treatment rules that are dominated. Thus the planner's immediate problem is to determine the treatment rules that are dominated given the available information.

To grasp the essence of the planner's problem, it suffices to consider the simplest non-trivial setting; that in which treatments, outcomes, and covariates are all binary. Thus I henceforth suppose that there are two treatments, say $t = 0$ and $t = 1$. The outcome $y(t)$ is binary, taking the values $y(t) = 0$ and $y(t) = 1$; hence $E[y(t)|x] = P[y(t) = 1|x]$. The covariate $x$ is also binary, taking the values $x = a$ and $x = b$.

There are four feasible treatment rules in this setting. These rules and their mean outcomes are

Treatment Rule $\tau(0, 0)$: All persons receive treatment 0. The mean outcome is $M(0, 0) \equiv P[y(0) = 1]$.

Treatment Rule $\tau(1, 1)$: All persons receive treatment 1. The mean outcome is $M(1, 1) \equiv P[y(1) = 1]$.

Treatment Rule $\tau(0, 1)$: Persons with $x = a$ receive treatment 0 and persons with $x = b$ receive treatment 1. The mean outcome is

$$M(0, 1) \equiv P[y(0) = 1|x = a] \cdot P(x = a)$$
$$+ P[y(1) = 1|x = b] \cdot P(x = b).$$

Treatment Rule $\tau(1, 0)$: Persons with $x = a$ receive treatment 1 and persons with $x = b$ receive treatment 0. The mean outcome is

$$M(1, 0) \equiv P[y(1) = 1|x = a] \cdot P(x = a)$$
$$+ P[y(0) = 1|x = b] \cdot P(x = b).$$

## 3.2. THE DOMINATED TREATMENT RULES

Which of the four feasible treatment rules are dominated? The experiment reveals $M(0, 0)$ and $M(1, 1)$. Thus rule $\tau(0, 0)$ is dominated if $M(0, 0) < M(1, 1)$ and rule $\tau(1, 1)$ is dominated if $M(1, 1) < M(0, 0)$. The planner is indifferent between these two rules if $M(0, 0) = M(1, 1)$.

The experimental data do not reveal $M(0, 1)$ and $M(1, 0)$. However, Manski (1997a, Proposition 7) shows that the experiment and the planner's knowledge of the covariate distribution imply sharp bounds on these quantities. The derivation begins from the fact that

$$(10a) \quad P[y(0) = 1] = P[y(0) = 1 | x = a] \cdot P(x = a)$$
$$+ P[y(0) = 1 | x = b] \cdot P(x = b)$$
$$(10b) \quad P[y(1) = 1] = P[y(1) = 1 | x = a] \cdot P(x = a)$$
$$+ P[y(1) = 1 | x = b] \cdot P(x = b).$$

Consider (10a). The planner knows $P[y(0) = 1]$ and $P(x)$. The unknowns $P[y(0) = 1 | x = a]$ and $P[y(0) = 1 | x = b]$ both lie in the interval $[0, 1]$. Hence (10a) yields informative sharp bounds on $P[y(0) = 1 | x = a]$ and $P[y(0) = 1 | x = b]$. Similarly, (11b) yields bounds on $P[y(1) = 1 | x = a]$ and $P[y(1) = 1 | x = b]$. The sharp bounds on $M(0,1)$ and $M(1, 0)$ then follow immediately. These turn out to be

$$(11a) \quad \max \{0, P[y(1) = 1] - P(x = a)\}$$
$$+ \max \{0, P[y(0) = 1] - P(x = b)\}$$
$$\leq M(0, 1)$$
$$\leq \min \{P(x = b), P[y(1) = 1]\}$$
$$+ \min \{P(x = a), P[y(0) = 1]\}$$
$$(11b) \quad \max \{0, P[y(1) = 1] - P(x = b)\}$$
$$+ \max \{0, P[y(0) = 1] - P(x = a)\}$$
$$\leq M(1, 0)$$
$$\leq \min \{P(x = a), P[y(1) = 1]\}$$
$$+ \min \{P(x = b), P[y(0) = 1]\}.$$

It might have been conjectured that $M(0, 1)$ and $M(1, 0)$ must lie in the interval $[M(0, 0), M(1, 1)]$. This is correct if treatment 0 is inferior to treatment 1 in subpopulations $\{x = a\}$ and $\{x = b\}$; that is, if $P[y(0) = 1 | x = a] \leq P[y(1) = 1 | x = a]$ and $P[y(0) = 1 | x = b] \leq P[y(1) = 1 | x = b]$. However the conjecture is not correct if the ordering of the mean outcomes differs across the two subpopulations. It is

this possibility that gives rise to the surprisingly complex bounds on $M(0, 1)$ and $M(1, 0)$ reported in (11).

The form of the bounds depends on the ordering of $P[y(0) = 1]$, $P[y(1) = 1]$, $P(x = a)$, and $P(x = b)$. Henceforth I assume without loss of generality that $P[y(0) = 1] \leq P[y(1) = 1]$ and $P(x = a) \leq P(x = b)$. Then there are six distinct orderings to be considered. For each one, application of (11a) and (11b) yields the bounds on $M(0, 1)$ and $M(1, 0)$. These bounds determine which treatment rules are dominated. The results follow: (Rule $\tau(0, 0)$ is dominated if $P[y(0) = 1] < P[y(1) = 1]$. I do not repeat this below.)

<u>Case 1</u>: $P[y(0) = 1] \leq P[y(1) = 1] \leq P(x = a) \leq P(x = b)$
$$0 \leq M(0, 1) \leq P[y(1) = 1] + P[y(0) = 1].$$
$$0 \leq M(1, 0) \leq P[y(1) = 1] + P[y(0) = 1].$$
Rules $\tau(0, 1)$, $\tau(1, 0)$, and $\tau(1, 1)$ are undominated.

<u>Case 2</u>: $P[y(0) = 1] \leq P(x = a) \leq P[y(1) = 1] \leq P(x = b)$
$$P[y(1) = 1] - P(x = a) \leq M(0, 1)$$
$$\leq P[y(1) = 1] + P[y(0) = 1].$$
$$0 \leq M(1, 0) \leq P(x = a) + P[y(0) = 1].$$
Rules $\tau(0, 1)$ and $\tau(1, 1)$ are undominated. Rule $\tau(1, 0)$ is dominated if $P(x = a) + P[y(0) = 1] < P[y(1) = 1]$.

<u>Case 3</u>: $P[y(0) = 1] \leq P(x = a) \leq P(x = b) \leq P[y(1) = 1]$
$$P[y(1) = 1] - P(x = a) \leq M(0, 1)$$
$$\leq P(x = b) + P[y(0) = 1].$$
$$P[y(1) = 1] - P(x = b) \leq M(1, 0)$$
$$\leq P(x = a) + P[y(0) = 1].$$
Rule $\tau(1, 1)$ is undominated. Rule $\tau(0, 1)$ is dominated if $P(x = b) + P[y(0) = 1] < P[y(1) = 1]$. Rule $\tau(1, 0)$ is dominated if $P(x = a) + P[y(0) = 1] < P[y(1) = 1]$.

<u>Case 4</u>: $P(x = a) \leq P[y(0) = 1] \leq P[y(1) = 1] \leq P(x = b)$
$$P[y(1) = 1] - P(x = a) \leq M(0, 1)$$
$$\leq P[y(1) = 1] + P(x = a).$$
$$P[y(0) = 1] - P(x = a) \leq M(1, 0)$$
$$\leq P(x = a) + P[y(0) = 1].$$
Rules $\tau(1, 1)$ and $\tau(0, 1)$ are undominated. Rule $\tau(1, 0)$ is dominated if $P(x = a) + P[y(0) = 1] < P[y(1) = 1]$.

Case 5: $P(x = a) \leq P[y(0) = 1] \leq P(x = b) \leq P[y(1) = 1]$

$P[y(1) = 1] - P(x = a) \leq M(0, 1) \leq 1.$

$P[y(1) = 1] + P[y(0) = 1] - 1 \leq M(1, 0)$

$\leq P(x = a) + P[y(0) = 1].$

Rules $\tau(1, 1)$ and $\tau(0, 1)$ are undominated. Rule $\tau(1, 0)$ is dominated if $P(x = a) + P[y(0) = 1] < P[y(1) = 1]$.

Case 6: $P(x = a) \leq P(x = b) \leq P[y(0) = 1] \leq P[y(1) = 1]$

$P[y(1) = 1] + P[y(0) = 1] - 1 \leq M(0, 1) \leq 1.$

$P[y(1) = 1] + P[y(0) = 1] - 1 \leq M(1, 0) \leq 1.$

Rules $\tau(0, 1)$, $\tau(1, 0)$, and $\tau(1, 1)$ are undominated.

Cases 1 through 6 show that as many as three or as few as zero treatment rules are dominated, depending on the empirical values of $P[y(0) = 1]$, $P[y(1) = 1]$, $P(x = a)$, and $P(x = b)$. The one constancy is that rule $\tau(1, 1)$ is always undominated; indeed, $\tau(1, 1)$ is always the maximin rule. I would emphasize that $\tau(1, 1)$ need not be the optimal rule. The fact that $\tau(1, 1)$ is always undominated simply means that, given the available information, the planner cannot reject the hypothesis that $\tau(1, 1)$ is the optimal rule.

# 4. AN EMPIRICAL ILLUSTRATION: THE PERRY PRESCHOOL PROJECT

An empirical illustration helps to see the range of possibilities. Beginning in 1962, the Perry Preschool Project provided intensive educational and social services to a random sample of low-income black children in Ypsilanti, Michigan. The project investigators also drew a second random sample of such children, but provided them with no special services. Subsequently, it was found that 67 percent of the treatment group and 49 percent of the control group were high-school graduates by age 19 (see Berrueta-Clement *et al.* (1984)).

Let $t = 1$ denote the Perry Preschool treatment and $t = 0$ denote the "no special services" control treatment. Let $y(t) = 1$ if a child receiving treatment $t$ is a high school graduate by age 19 and $y(t) = 0$ otherwise. Abstracting from sampling variability and ignoring some attrition from the experiment, the outcome data reveal that $P[y(0) = 1] = 0.49$ and $P[y(1) = 1] = 0.67$.

Consider the situation of a planner, perhaps a social worker, who is charged with making preschool treatment choices for low-income black children in Ypsilanti and whose objective is to maximize the high school graduation rate. The planner can assign each child to the Perry Preschool treatment or not. Suppose that the planner observes a binary covariate that describes each member of the population. For the sake of concreteness, let the covariate indicate the child's family status, with $x = a$ if the child has an intact two-parent family and $x = b$ otherwise.

The available outcome data reveal that treatment rule $\tau(0, 0)$, wherein no children receive the Perry Preschool treatment, is dominated by rule $\tau(1, 1)$, wherein all children receive preschooling. The conclusions that the planner can draw about rules $\tau(0, 1)$ and $\tau(1, 0)$ depend on the covariate distribution $P(x)$.

Suppose that half of all children have intact families, so $P(x = a) = P(x = b) = 0.5$. Then Case 3 of Section 3.2 holds. The bounds on mean outcomes under rules $\tau(0, 1)$ and $\tau(1, 0)$ are

$0.17 \leq M(0, 1) \leq 0.99 \qquad 0.17 \leq M(1, 0) \leq 0.99.$

These bounds imply that rules $\tau(0, 1)$ and $\tau(1, 0)$, which reverse one another's treatment assignments, have an enormously wide range of potential consequences for high school graduation. The best case for $\tau(0, 1)$ and the worst for $\tau(1, 0)$ both occur if the (unknown) graduation probabilities conditional on covariates are

$P[y(0) = 1 | x = a] = 0.98 \quad P[y(1) = 1 | x = a] = 0.34$

$P[y(0) = 1 | x = b] = 0 \qquad P[y(1) = 1 | x = b] = 1.$

These graduation probabilities, which yield $M(0, 1) = 0.99$ and $M(1, 0) = 0.17$, are consistent with the experimental evidence that $P[y(0) = 1] = 0.49$ and $P[y(1) = 1] = 0.67$. They describe a possible world in which preschooling is necessary and sufficient for children in non-intact families to complete high school, but substantially hurts the graduation prospects of children in intact families. There is another possible world with the reverse graduation probabilities, one in which $M(0, 1) = 0.17$ and $M(1, 0) = 0.99$. Hence rules $\tau(0, 1)$, $\tau(1, 0)$, and $\tau(1, 1)$ are all undominated.

The planner faces a much less ambiguous choice problem if most children have non-intact families. Suppose that $P(x = a) = 0.1$ and $P(x = b) = 0.9$. Then Case 4 of Section 3.2 holds. The bounds on mean outcomes under rules $\tau(0, 1)$ and $\tau(1, 0)$ are

$0.57 \leq M(0, 1) \leq 0.77 \qquad 0.39 \leq M(1, 0) \leq 0.59.$

These bounds are much narrower than those obtained when half of all children have non-intact families. The upper bound on $M(1, 0)$ is 0.59, which is less than the known value of $M(1, 1)$, namely 0.67. Hence treatment rule $\tau(1, 0)$ is dominated. Recall that rule $\tau(0, 0)$ is also dominated.

Thus, although the planner does not observe graduation probabilities conditional on covariates, he can nevertheless conclude that the 90 percent of children who have non-intact families should receive preschooling. The only ambiguity about treatment choice concerns the 10 percent of children who have intact families. Treatment rules τ(0, 1) and τ(1, 1) are undominated. Thus, given the information available, the planner cannot determine whether these children should or should not receive preschooling.

## 5. CONCLUSION

In principle, the informational problem analyzed in this paper would disappear if researchers performing randomized experiments would collect extensive covariate data for their subjects and report outcome distributions conditional on these covariates. There seem to be two reasons why researchers commonly report experimental outcomes without much covariate information.

First, researchers often seem to assume that treatment effects are constant across the population, or at least that they do not vary in sign across different subpopulations. (I say "seem to" because the assumption is commonly implicit, not explicit.) Given two treatments t = 0 and t = 1, suppose one knows a priori that there are only these two possibilities: either {E[y(1)|x] ≥ E[y(0)|x], all x ∈ X} or {E[y(1)|x] ≤ E[y(0)|x], all x ∈ X}. Then collection of data on the covariate x is not necessary to determine an optimal treatment rule. It suffices to learn if E[y(1)] exceeds E[y(0)].

Second, there is the matter of sampling variability. Researchers often perform randomized experiments with samples of subjects that are large enough to yield statistically precise findings for unconditional mean outcomes but not large enough to yield precise findings for mean outcomes conditional on covariates. Findings conditional on covariates commonly go unreported if they do not meet conventional criteria for statistical precision.

Whatever the reasons are for the dearth of covariate data in reports of randomized experiments, the implications for treatment choice are clear. A planner who observes covariates that describe each member of the population can choose treatment rules that differentiate among persons with different covariate values. Lacking knowledge of mean outcomes conditional on the observed covariates, the planner generically faces a problem of treatment choice under ambiguity. Except in special circumstances (e.g., the subcase of Case 3 in which three of the four feasible rules are dominated), the planner cannot determine whether using the observed covariates to choose treatments improves or degrades the quality of decision making. This places the planner in an interesting, if disquieting, predicament. I cannot say how actual planners — physicians, social workers, etc. — cope with this predicament in practice.

## Acknowledgments

The research reported here is supported in part by National Science Foundation Grant SBR-9722846. I am grateful to John Pepper and two anonymous reviewers for helpful comments.

## References

Berrueta-Clement, J., L. Schweinhart, W. Barnett, A. Epstein, and D. Weikart. 1984. Changed Lives: The Effects of the Perry Preschool Program on Youths Through Age 19, Ypsilanti, Michigan: High/Scope Press.

Manski, C. 1989. "Anatomy of the Selection Problem." Journal of Human Resources, 24: 343-360.

Manski, C. 1990. "Nonparametric Bounds on Treatment Effects." American Economic Review Papers and Proceedings 80: 319-323.

Manski, C. 1993. "Identification Problems in the Social Sciences." pp. 1-56 in Sociological Methodology 1993, edited by Peter V. Marsden, Cambridge, MA: Blackwell Publishers.

Manski, C. 1994. "The Selection Problem." pp. 143-170 in Advances in Econometrics, Sixth World Congress, edited by Christopher Sims. Cambridge, England: Cambridge University Press.

Manski, C. 1995. Identification Problems in the Social Sciences. Cambridge, Mass.: Harvard University Press.

Manski, C. 1996. "Learning about Treatment Effects from Experiments with Random Assignment of Treatments." Journal of Human Resources 31: 707-733.

Manski, C. 1997a. "The Mixing Problem in Programme Evaluation." Review of Economic Studies 64: 537-553.

Manski, C. 1997b. "Monotone Treatment Response." Econometrica 65: 1311-1334.

Manski, C. 1998. "Treatment Under Ambiguity." in S. Durlauf, J. Geanakoplos, and J. Traub (editors), Fundamental Limits to Knowledge in Economics, Reading, Mass.: Addison-Wesley, forthcoming.